\documentclass[]{article}
\usepackage{graphicx}
\usepackage{amsmath, amssymb, latexsym}
\usepackage{fullpage, subfigure, comment}
\usepackage{algorithm}
\usepackage{algpseudocode}

\newtheorem{example}{Example}
\newtheorem{claim}{Claim}
\newtheorem{thm}{Theorem}
\newtheorem{defn}{Definition}

\title{Recognizing and generating unswitchable graphs}
\author{Asish Mukhopadhyay, Daniel John and Srivatsan Vasudevan \\
	School of Computer Science \\
	Univesity of Windsor\\
	Ontario, Canada}

\begin{document}
	
\maketitle
	
\begin{abstract}
		In this paper, we show that unswitchable graphs are a proper subclass of  split graphs, 
		and exploit this fact to propose efficient algorithms for their recognition and generation. 
\end{abstract}

\section{Introduction}

Let $G = (V, E)$ be a simple graph on the vertex set $V = \{v_1, v_2, \ldots, v_n\}$. Let $d_i$ be the degree of $v_i$. Assume 
without loss of generality that $n- 1 \geq d_1 \geq d_2 \geq \ldots \geq d_n \geq 0$. There may exist many other graphs $G$ with the 
same degree sequence, making it a many-to-one mapping.\\

\begin{figure}[h!]
	\centering
	\includegraphics[scale=0.6]{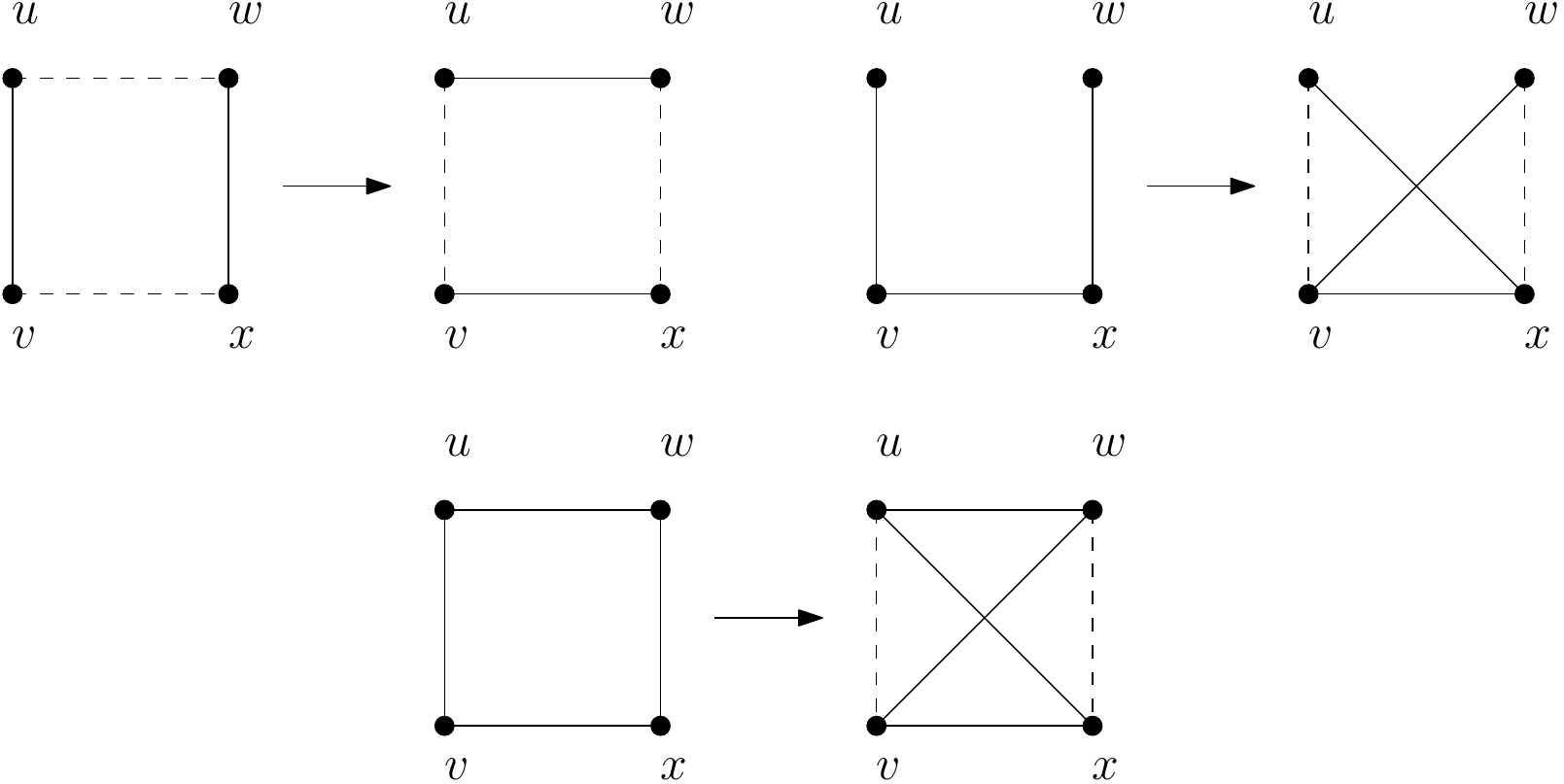}
	\caption{\textit{Graph $G'$ obtained from $G$ by a 2-switch}}
	\label{twoSwitch}
\end{figure}

Let the edges $\{u, v\}$,  $\{w, x\}$ of  $G$ be independent (this means that the edges do not have an end-point in common). For each of the three ways the edges can be independent, we can obtain another graph $G'$ with the same degree 
sequence by means of a 2-switch as shown in Figure~\ref{twoSwitch}, in pairs from left to right and top to bottom, where the dashed lines show the replacement edges. 

A graph $G$ is said to be \textit{unswitchable} if it cannot be reduced to another graph $H$ with the same degree sequence by 
edge-switching.
In this paper, we propose an algorithm for recognizing unswitchable graphs that exploits the relationship of this class of
graphs to the class of split-graphs. \\

To motivate the significance of the concept of edge-switching we dicuss an application in the next section.

\section{An application of edge-switching}

Given a graph $G$, it is easy to obtain its degree sequence $d$. However, given $d$ with the $d_i$'s in non-increasing
order, the question whether there exists a graph whose degree sequence is $d$ has spawned a lot of research.  We begin with the following definition.

\begin{defn}
	The sequence $d$ is graphical if  there exists a graph $G$ such $d_{G}(v_i) = d_i$ for
	$i = 1, \ldots, n$, where $d_{G}(v_i)$ denotes the degree of the vertex $v_i$ in $G$. 
\end{defn}

Both Hakimi~\cite{hakimi1962} and Havel~\cite{havel1955} are credited with the following result.

\begin{thm}
	Let $n \geq 2$ and $d_ 1 \geq 1$. The sequence $d$ is graphical if and only if the sequence $d_2 -1, d_3 -1, \ldots, d_{d_1+1} - 1, d_{d_1 + 2}, d_{d_1 + 3}, \ldots, d_n$, arranged in nonincreasing order, is graphical.	
\end{thm}

To prove this result we need a definition and prove two other results. \\ 

If a graph $H$ can be obtained from a graph $G$ by a finite sequence of 2-switches we indicate this 
reduction by the notation $G \stackrel{2s}{\Longrightarrow} H$. Berge~\cite{berge1973} proved that:

\begin{thm}\label{thmx}
	Two graphs $G$ and $H$ on a common vertex set $V$ satisfy
	$d_G(v) = d_H(v)$ for all $v \in V$ if and only if $G \stackrel{2s}{\Longrightarrow} H$.
\end{thm}

We will invoke this result when we introduce unswitchable graphs later on. 
To prove Theorem~\ref{thmx}, we first prove the following result, given a non-increasing degree sequence $d$
as above.

\begin{thm}\label{thmy}
	If $G$ be a graph on $n$ vertices such that $d_G(v_i) = d_i$, then there exists a graph $G'$ such that $G \stackrel{2s}{\Longrightarrow}G'$ with $N_{G'}(v_1) = \{ v_2, \ldots, v_{d_1 + 1}\}$.	
\end{thm}

{\bf Proof:} Let $d = \Delta(G) (= d_1)$ be the maximum vertex degree of $G$. Assume there 
exists a $v_i$  such that $\{v_1, v_i\} \notin E$ for  
$i$ in the range $[2, d+1]$.  Instead, there is an index $j \geq d+2$ such that  
$\{v_1, v_j\} \in E$. Again, as $j > i$, according to our assumption on the degree sequence,
$d_j \leq d_i$. If $V_i$ and $V_j$ are the subsets of vertices of $V$  that $v_i$ and $v_j$ are connected to respectively, $V_i - V_j \neq \emptyset$. Hence there exists $t$ such that $\{v_i, v_t\} \in E$, but 
$\{v_j, v_t\} \notin E$. Thus we can make a 2-switch so that $v_1$ is adjacent to $v_i$. We repeat this till all the vertices adjacent to $v_1$ have indices in the range $[2, d+1]$. \hfill$\blacksquare$\\

Berge's theorem is easily proved by induction on the number of vertices of the graphs $G$ and $H$. 
The condition is sufficient as $G \stackrel{2s}{\Longrightarrow} H$ means that the vertex degrees are preserved. Conversely, by applying Theorem~\ref{thmy} to each of the graphs $G$ and $H$ we can find a vertex $v$ such that in graphs $G'$ and $H'$ respectively where 
$G \stackrel{2s}{\Longrightarrow} G'$ and $H \stackrel{2s}{\Longrightarrow} H'$, the neighborhood of $v$ is identical. Now the reduced graphs $G' - v$ and $H' - v$ have the same
vertex degrees and by the induction hypothesis $G' - v  \stackrel{2s}{\Longrightarrow} H' -v$. 
Consequently, $G' \stackrel{2s}{\Longrightarrow} H'$. Combining this with the fact that 
$H' \stackrel{2s}{\Longrightarrow} H$ by a sequence of reverse 2-switches, the necessity is proved.\\

Here's is an interesting application of Berge's result. Consider the example below 
where we want to reduce graph $G$ to graph $H$ by 2-switches so that $v_1$ is adjacent 
to $v_2$ and $v_3$. We achieve this by switching the pair of edges $\{v_2, v_3\}$,  $\{v_1, v_4\}$ with the non-existing pair of edges $\{v_1, v_3\}$,  $\{v_2, v_4\}$. \\

\begin{figure}[h!]
	\centering
	\includegraphics[scale=0.6]{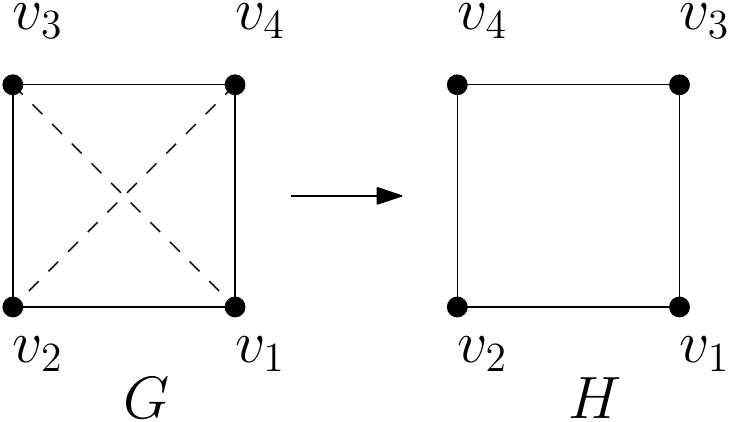}
	\caption{\textit{Graph $H$ obtained from $G$ by a 2-switch}}
	\label{twoSwitchExample}
\end{figure}

Now, we can prove Theorem 1.\\

{\bf Proof:} Consider the \textit{if} direction. Let $G$ be a graph
on $n-1$ vertices with the degree sequence: \\

$\langle d_G(v_2) = d_2 - 1, d_G(v_3) = d_3 - 1, \ldots, d_G(v_{d_1 + 1})= d_{d_1 + 1} - 1, d_G(v_{d_1 + 2})= d_{d_1 + 2} , d_G(v_{d_1 + 3})= d_{d_1 + 3} , \ldots , d_G(v_n) = d_n \rangle$\\

Add a new vertex $v_1$ and the edges $\{v_1, v_i\}$ for all $i \in [2, d_{d_1 + 1}]$ . Then in the
new graph $H$, $d_H(v_1) = d_1$ , and $d_H(v_i) = d_i$ for all $i \geq 2$. \\

For the  \textit{only if} direction, assume $d_G(v_i) = d_i$. By the Lemma proved earlier
and Berge's result, we can assume that
$N_G(v_1) = \{ v_2, \ldots, v_{d_1 + 1 }\}$. 
But now the degree sequence of $G - v_1$ is as above.  

\begin{example}
The sequence $\langle 4, 4, 4, 3, 2, 1 \rangle$ is graphical since 
the following sequence of reduced sequences are each graphical: $\langle 3, 3, 2, 1, 1 \rangle$, 
$\langle 2, 1, 1, 0 \rangle$ (this is obtained by a reordering of $\langle 2, 1, 0, 1 \rangle$,
obtained from the previous sequence), $\langle 0, 0, 0 \rangle$. The last sequence corresponds to an 
empty graph, and the graph corresponding to the initial sequence is easily constructed.
\end{example}

It should be pointed out that Hakimi's algorithm will work if the sequence element that we choose to saturate 
is any element of the sequence. If its degree is $d_i$, we reduce the $d_i$ \textit{highest degree elements} by 1. This observation is due to Kleitman and Wang ~\cite{DBLP:journals/dm/KleitmanW73}. 

\section{Split graphs}\label{splitGraphs}
\begin{figure}[h!]
	\centering
	\includegraphics[scale=0.6]{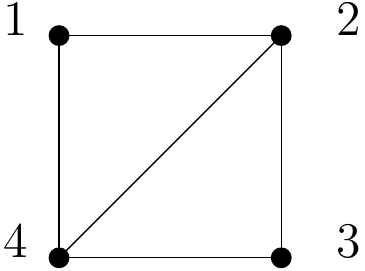}
	\caption{\textit{A split graph}}
	\label{splitGraph}
\end{figure}

A graph $G$ is said to be a split graph if there exists a disjoint partition of its vertex set $V$ into a complete induced subgraph on $V_2$ vertices and an 
independent set of $V_1$ vertices. Fig.~\ref{splitGraph} shows an example of a split graph where the induced subgraph on the vertices $\{2,4\}$
is complete and the subset of vertices $\{1,3\}$ form an independent set. 

The partition of the graph into a complete graph and an independent set is not unique. For the example split graph, $\{1, 2, 4\}$ and 
$\{3\}$ is another partition into a (maximal) complete graph and an independent set. \\

\begin{figure}[h!]
	\centering
	\includegraphics[scale=0.6]{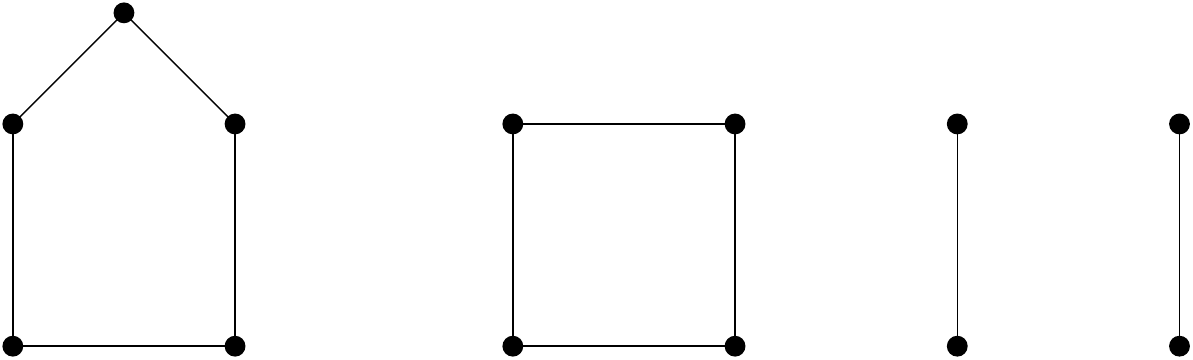}
	\caption{\textit{Forbidden subgraphs of a split graph}}
	\label{forbiddenSubgraphs}
\end{figure}

There are other characterizations of clique graphs. For example, this: A graph $G$ is a split graph iff it does not contain any of the  graphs of Figure~\ref{forbiddenSubgraphs} as induced subgraphs. \\

There is yet another characterization of a split graph in terms of the degrees of its vertices \cite{Hammer1981TheSO}. Let $d = (d_1, d_2, \ldots, d_n)$ be the sequence of degrees of its vertices, with $n-1 \geq d_1  \geq d_2 \geq  d_3 \geq  \ldots \geq d_n \geq 0$. Let $m$ be the maximum index $i$ for which $d_i \geq i-1$. Call it the split index.\\

Then $G$ is a split graph iff: 

\begin{equation}\label{EG}
\Sigma_{i=1}^{m} d_i = m(m-1) + \Sigma_{i=m+1}^{n} d_i
\end{equation}

Thus for the example split graph of Fig.~\ref{splitGraph}, we have $d = (3, 3, 2, 2)$, $m = 3$ and both sides of Eqn.(\ref{EG}) evaluate to 8. \\

This last characterization forms the basis for an easy recognition algorithm for split graphs. From the degree sequence, 
find the split index $m$, going left to right in the degree sequence. Construct a complete graph on $m$ vertices with degree $d_m$. 
The remaining vertices form an independent set, which are now joined to the clique vertices to saturate their degrees and
the residual degrees of the clique vertices. \\

The forbidden subgraph characterization is 
of interest to us. If a split graph has a 4-cycle or its complement as an induced subgraph then it is switchable. The question
is: Are all split-graphs switchable ? We explore this matter in the next section.

\section{Unswitchable graphs}

A $P_4$ is a chordless path on 4 vertices of $G$, while a $C_4$ is a 4-cycle and a $2K_2$ (the complement of a 4-cycle) is a subgraph with 2 disjoint edges of $G$.\\

Clearly, an unswitchable graph $G$ cannot have a $P_4$, a $C_4$ or a $2K_2$ as an induced subgraph on 4 vertices. Since 
no switching is possible, we cannot use 2-switches to transform a given graph $G$ to a graph $G'$ with the 
same degree sequence. \\

Extrapolating from the forbidden induced subgraph characterization of unswitchable graphs, Eggleton proposed the 
following constructive charaterization of unswitchable graphs. 

\begin{thm}\cite{eggleton1975}
	For any positive integer $n$, let $\{S_i: 1 \leq i \leq 2n\}$ be a family of pairwise disjoint 
	finite (possibly empty) sets, with union $V$. Let $G$ with a vertex set $V$, such that any two distinct 
	vertices $a \in S_i$ and $b \in S_j$, with $i \leq j$, are adjacent in $G$ just if $i + n < j$ or $i > n$. 
	Then $G$ is unswitcahable; moreover every unswitchable graph is obtained by this construction. 
\end{thm}

{\bf Proof:} (Ours) We show that the graph constructed cannot have any of the graphs $P_4$, $C_4$ or $2K_2$ as an induced subgraph. We argue the case of
$C_4$. Let the labels of the vertices of $C_4$ be $a, b, c, d$ in cyclic order. Since $a$ and $c$ are not connected both cannot be in sets with indices greater than $n$. Let $a$ be in a set $S_i$ with index $i \leq n$. Since $a$ is joined to both $b$ and $d$ they are in sets $S_j$ and $S_k$ with indices greater than $i + n$. Thus $b$ and $d$ must be connected. This contradicts the assumption that the induced graph on $a, b, c, d$ is a $C_4$. \\

Similar argumemts can be made for the non-existence of $2K_2$ and $P_4$ as induced subgraphs. \\

Now for the second half of the theorem. Let $G$ be a given unswitchable graph. Since it is a split graph, 
let there be $m$ edges connecting a vertex of the independent set with a vertex of the clique. If $\{u, v\}$ 
is one such edge, let $u \in S_{i_1}$ and $v \in S_{j_1}$. Then we must have $j_1 - i_1 > n$. Thus we have 
$m$ such inequalities corresponding to the $m$ edges.\\

Further, $j_1 > n$ and $i_1 \leq n$ for each pair of indices corresponding to the $m$ edges. This means that we have to choose
$m$ pairs of points in the polygonal region in the $x-y$ plane bounded by the lines $x-y > n$, $x > n$ and 
$y \leq n$. \\

\begin{figure}[h!]
	\centering
	\includegraphics[scale=0.6]{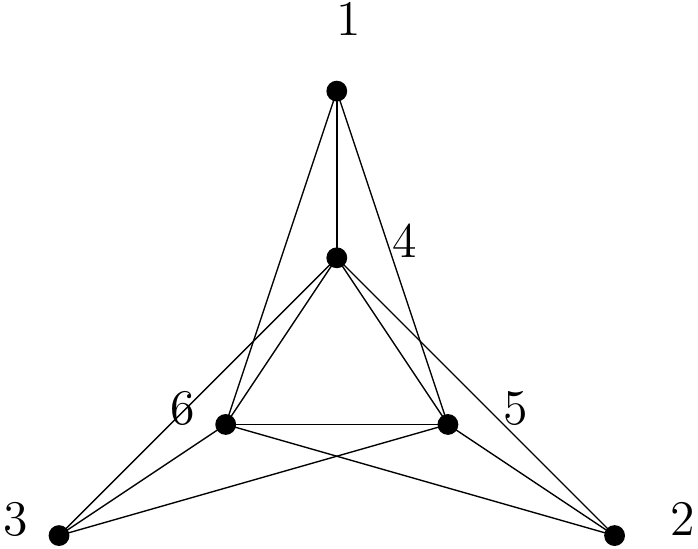}
	\caption{\textit{A non-switchable graph $G_1$}}
	\label{nsGraph}
\end{figure}

\begin{figure}[h!]
	\centering
	\includegraphics[scale=0.6]{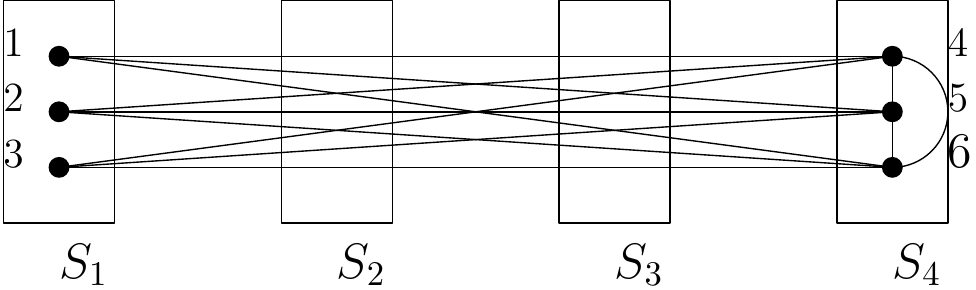}
	\caption{\textit{Set distribution of the vertices of $G_1$}}
	\label{vertexDistribution}
\end{figure}

We choose a minimum $n$ such that $m$ pairs of points can be found in this polygonal region. For 
the unswitchable graph of Figure~\ref{nsGraph} a distribution of its vertices among the sets $S_i$ is shown
in Figure~\ref{vertexDistribution} \hfill $\blacksquare$\\

Following the theorem, we constructed the unswitchable graph shown in Figure~\ref{nsGraph}, setting $n = 2$.\\

The sets $S_i$, the membership of the vertices 
in these sets and the mutual adjacencies of the vertices are shown in Figure~\ref{vertexDistribution}.\\

Here's another example, where we have gone in the opposite direction, setting $n = 2$ again and constructing the sets $S_i$, 
for $i = 1, 2, \ldots, 2n$, and adding edges between vertices in pairs of sets $S_i$ and $S_j$ for $i \leq j$, satisfying 
the other constraints on $i$ and $j$.  \\

\begin{figure}[h!]
	\centering
	\includegraphics[scale=0.6]{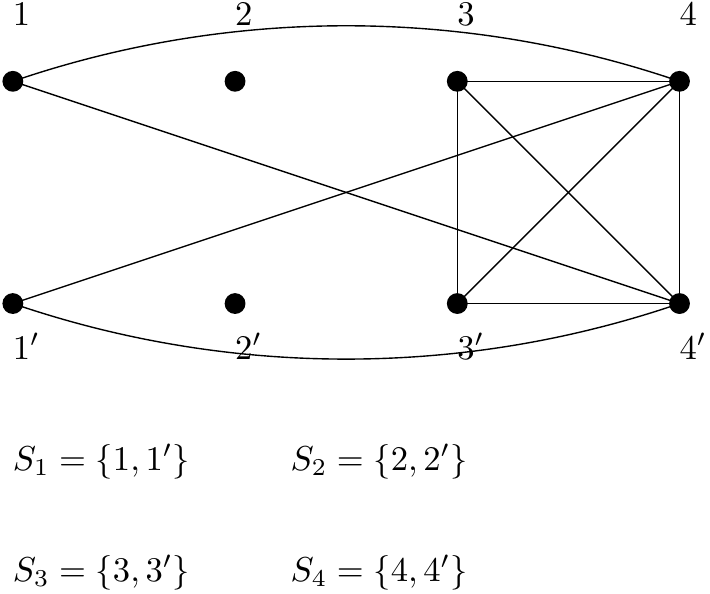}
	\caption{\textit{Distribution of the vertices among the sets $S_i$}}
	\label{vertexDistributionAnother}
\end{figure}

Both the graphs of Figure~\ref{nsGraph} and Figure~\ref{vertexDistributionAnother} are split-graphs. This leads us to 
speculate on what might be the relationship between these two graph classes: split-graphs and unswitchable graphs. \\

\begin{figure}[h!]
	\centering
	\includegraphics[scale=0.6]{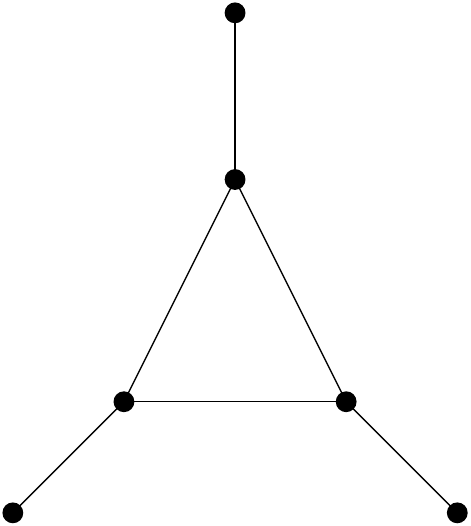}
	\caption{\textit{A split graph that is switchable}}
	\label{sButSplit}
\end{figure}


It appears that the class of split graphs has an overlap with the class of unswitchable graphs. 
As evidence we have the graphs of Figure~\ref{nsGraph} and Figure~\ref{vertexDistributionAnother} which are split graphs 
but not switchable. On the other hand the graph of Figure~\ref{sButSplit} is a split graph but  switchable 
as there exist several $P_4$'s as induced subgraphs. \\

\begin{figure}[h!]
	\centering
	\includegraphics[scale=0.6]{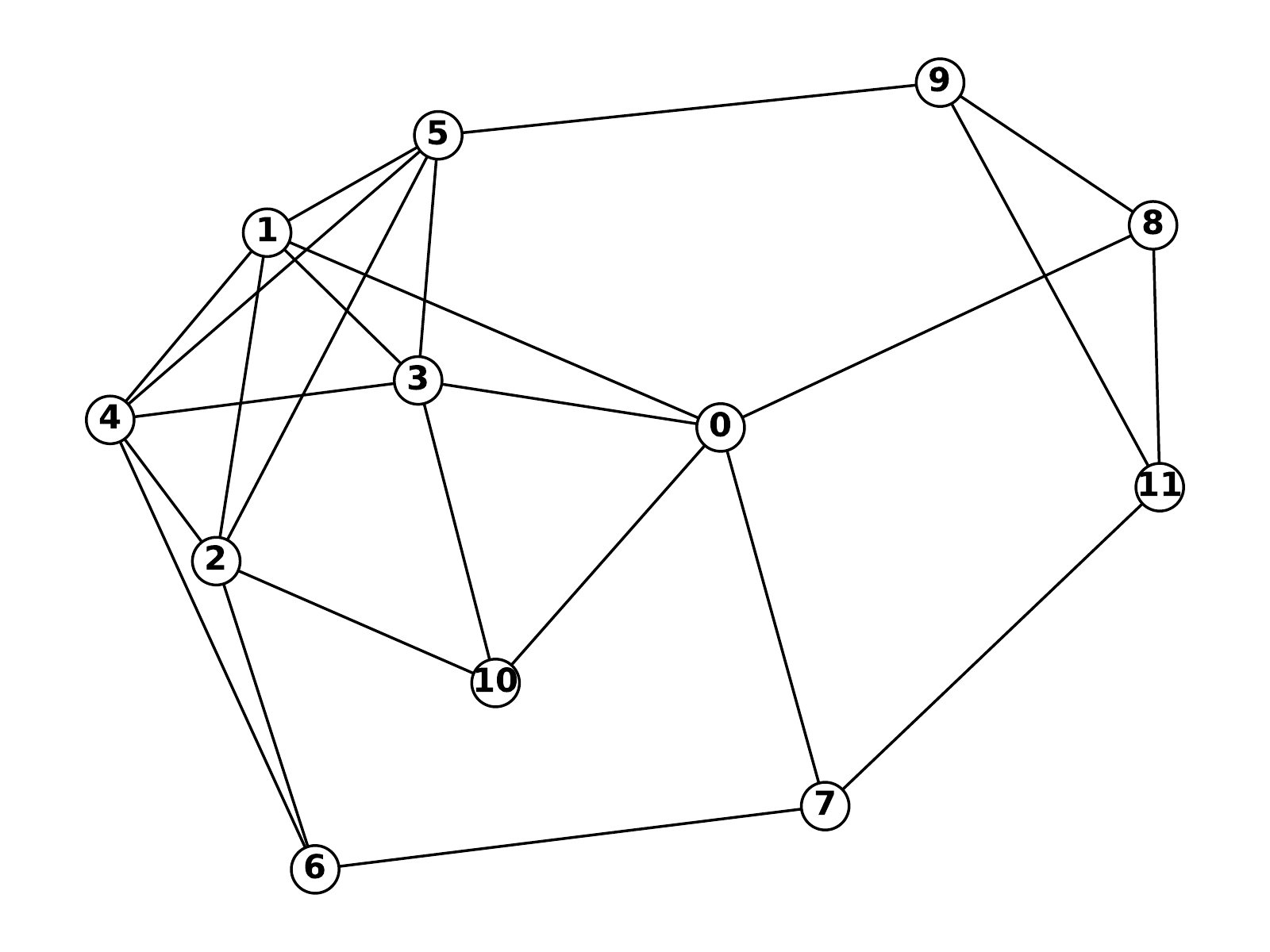}
	\caption{\textit{A graph that is switchable but not split }}
	\label{sButNotSplit}
\end{figure}

An interesting problem is to construct a switchable graph that is not a split graph.
Consider a graph $G$ consisting of two copies of the graph of Figure~\ref{nsGraph}. This graph is not a split graph but it is switchable. Indeed by running our implementation of Hakimi's 
algorithm on the degree sequence $d = (5, 5, 5, 5, 5, 5, 3, 3, 3, 3, 3, 3)$ we obtained the graph of Figure~\ref{sButNotSplit} as output. This is not a split graph as there is an induced 4-cycle on the vertex set $\{2, 3, 9, 10\}$ and is a switchable graph for the same reason.\\

The above considerations lead us to make the following claim.

\begin{claim}\label{claim}
Unswitchable graphs are a proper subclass of split graphs. 
\end{claim}

{\bf Proof:}
This is true since the graphs defined by Eggleton's result are all split graphs. The vertices in the sets with indices at most $n$ 
constitute an independent set and the ones with indices greater than $n$ form a complete graph. The inclusion is proper since we 
have found a split graph that is switchable (Figure~\ref{sButSplit}).  \hfill $\blacksquare$\\

In view of Claim~\ref{claim}, we can design a recognition algorithm for \textit{unswitchable graphs}. Given an input graph, we first run a 
recognition algorithm for split graphs (for example, the degree sequence based recognition algorithm mentioned in the 
previous section) and if the output is true, check that the graph does not have a $P_4$ as an induced 
subgraph. For this we proceed as follows.\\

\begin{figure}[h!]
	\centering
	\includegraphics[scale=0.7]{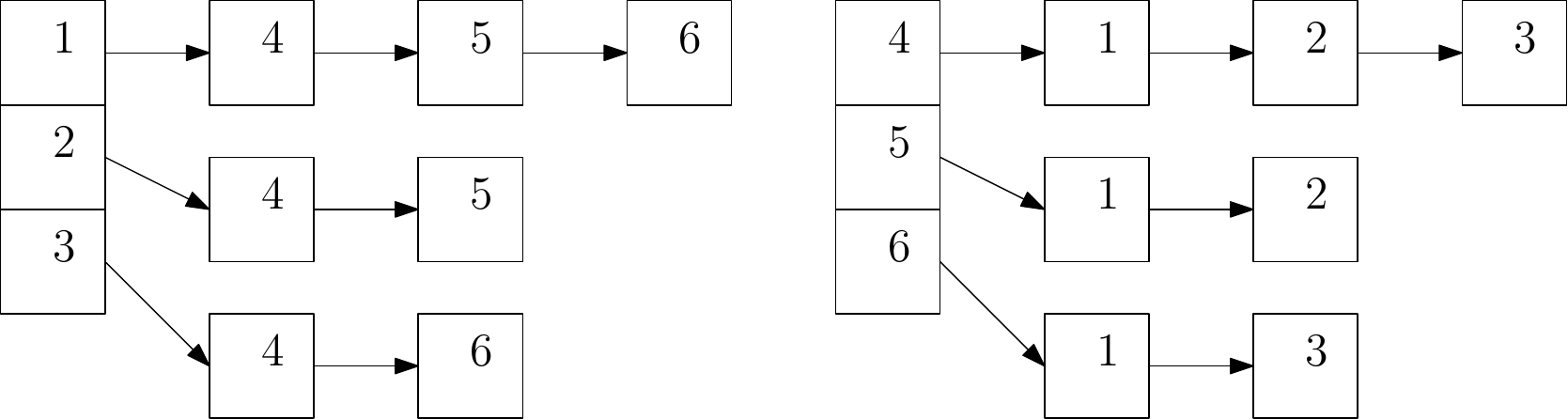}
	\caption{\textit{Adjacency Lists for the modified Graph of Fig.~\ref{nsGraph}}}
	\label{adjacencyLists}
\end{figure}

The recognition algorithm returns a split index $m$ as discussed in the Section~\ref{splitGraphs} so that the vertices
with degrees $d_{m+1} \geq d_{m+2} \geq \ldots \geq d_n$ constitute an independent set. 
Knowing this, from the adjacency list of the input graph, we find the adjacency list of each vertex of the independent set (Figure~\ref{adjacencyLists}). \\

We use this information to construct another adjacency list that gives for each vertex of the complete graph the vertices  of the independent set that are adjacent to it. \\

Now, for an edge $\{u, v\}$ of the clique we can find the sets of vertices $S_u$ and $S_v$  of the independent set that are adjacent to $u$ and $v$ respectively. If the set differences $S_u - S_v$ and $S_v - S_u$ are both nonempty then there exists a path $P_4$ 
betweeen $u$ and $v$, making the graph switchable. If there exists no clique edge $\{u, v\}$ for which this is true then the graph is unswitchable. \\

Consider the graph of Figure~\ref{nsGraph} without the edges 2-6 and 3-5. The adjacency list 
for the vertices of the independent set and the adjacency list for the vertices of the complete graph derived from it are shown in Figure~\ref{adjacencyLists}. \\

For the edge $u$-$v$ = 6-5, $S_u = \{1, 3\}$ and $S_v = \{1, 2\}$. The set differences are 
$\{2\}$ and $\{3\}$. Since these are both non-empty, there is a $P_4$ path: 3-6-5-2. This 
shows that the modified graph is switchable. \\

A formal description of the recognition algorithm is given below. The time complexity of the 
recognition of the algorithm is $O(n_1n_2^2)$, where $n_1$ and $n_2$ are respectively the sizes of the 
independent set and the clique set. Since $n_1$ and $n_2$ are both bounded by $n$, $O(n^3)$ is 
a more succinct description of the complexity of the algorithm. 

\begin{algorithm} 
	\caption{\textit{UnswitchableGraphRecognition}$(G)$}
	\label{alg:fulkersonRyserRealization}
    {\bf Input:} {The adjacency lists of the vertices of a graph $G$}\\
	{\bf Output:} {$G$ is switchable or not}
	\begin{algorithmic}[1]  
	\State Extract the degree sequence, $d = d_1 \geq d_2 \geq \ldots \geq d_n$ of $G$
    \State output $\leftarrow$  Run the recognition algorithm for a split-graph on $d$ 
	\If {(output = YES)} 
		\State Let $m$ be the split index
		\State Extract adjacency lists of the vertices with degrees $\geq d_{m+1}$
		\For {each edge $u-v$ of the complete graph on the vertices with degrees $\leq d_m$:}
			\State Compute the neighborhoods $S_u$ and $S_v$ of the end points in the independent set 
			\State Compute $S_u - S_v$ and $S_v - S_u$. 
			\If {$S_u - S_v$ and $S_v - S_u$ are disjoint and non-empty:}
				\State \Return ``G is switchable"
			\Else ~continue
			\EndIf
		\EndFor
		\State \Return ``G is unswitchable"
	\Else 
		\State \Return ``G is switchable"
	\EndIf
\end{algorithmic}
\end{algorithm}

\section{Generating an unswitchable graph}

The second half of the proof of Eggleton's theorem requires an unswitchable graph as input. We would also like to test the recognition algorithm of the previous section on instances of unswitchable graphs. Motivated by these applications, we consider the problem of generating an unswitchable graph on $n$ vertices by an independent method.\\

We first generate a split graph. Let $n$ be the number of vertices $V$ of the graph, obtained as (user) input. We partition $V$ into two disjoint non-empty 
subsets $V_1$ and $V_2$ of size $n_1$ and $n_2$ respectively. We assume that $n_2 \geq 2$ to avoid trivial cases. Construct a complete graph on the vertices of $V_2$. For each of the remaining $n_1$ vertices of $V_1$, 
choose a random integer $p$ in the range [0, $n_2$] and join the chosen vertex to a random subset of vertices of $V_2$ of size $p$.\\

We now proceed as in the algorithm for recognizing a split graph with a small change. 
For each pair of vertices $\{u, v\}$  in the independent set $V_1$, we determine the set of neighbors $S_u$ and $S_v$ in the  set of clique vertices $V_2$.
Compute the difference sets $S_u$-$S_v$ and $S_v$-$S_u$. If these are non-empty and disjoint, 
for each pair of  vertices $x$ and $y$ in the difference sets we have a $P_4$, defined by 
$u$-$x$-$y$-$v$. \\

A formal algorithm for generating these $P_4$'s is described below. If no induced subgraph isomorphic to a $P_4$ has been found, then we have an unswitchable graph. Othewise, we introduce new edges (chords) to eliminate the $P_4$'s. This in turn will 
generate new $P_4$'s formed by pairs of the newly introduced chords. Once again chords are introduced to eliminate the new $P_4$'s. We continue until only one new $P_4$ is generated.\\

\begin{algorithm}
	\caption{\textit{findP4s}}
	\begin{algorithmic}[1]
		\State \textbf{Input:} Adjacency list of graph $G$ and vertices of the independent set $V_1$
		\State \textbf{Output:} List of all $P_4$s in $G$
		\Procedure{find\_all\_P4s}{$adjacency\_list, V_1$}
		\State $ListP4 \leftarrow []$
		\For {each pair of vertices $v_1$ and $v_2$ in $V_1$ }
		\State $S_1 \leftarrow$ set of neighbours of $v_1$ read from $adj\_list$
		\State $S_2 \leftarrow$ set of neighbours of $v_2$ read from $adj\_list$
		\State $S_{12} \leftarrow S_1 - S_2$
		\State $S_{21} \leftarrow S_2 - S_1$
		\If {both $S_{12}$ and $S_{21}$ are non-empty}
		\For {each pair $a,b$ where $a \in S_{12}$ and $b \in S_{21}$ }
		\State Add $[v_1, a, b, v_2]$ to $ListP4$
		\EndFor
		\EndIf
		\EndFor
		\State \textbf{return} $ListP4$
		\EndProcedure
	\end{algorithmic}
\end{algorithm}

\begin{figure}[h!]
	\centering
	\includegraphics[scale=0.6]{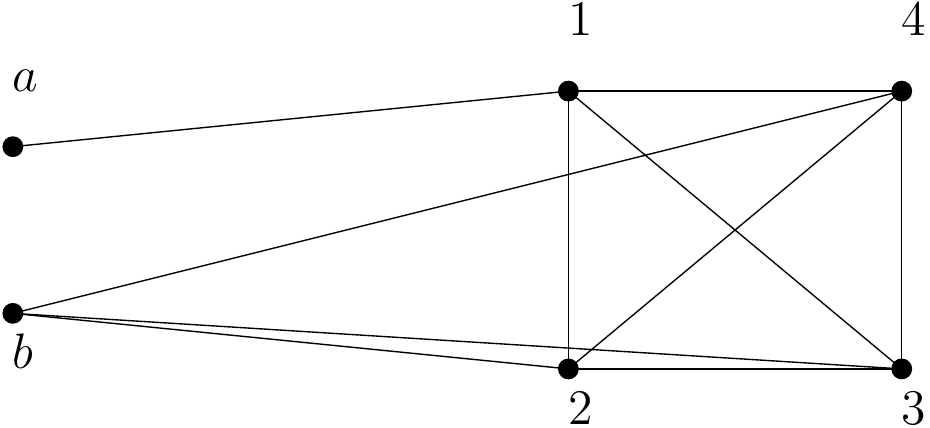}
	\caption{\textit{First step in generating an unswitchable graph}}
	\label{uggOne}
\end{figure}

Consider the example of Figure~\ref{uggOne}, where $V_1 = \{a,b\}$ and $V_2 = \{1, 2, 3, 4\}$. Apart from the edges of the 
clique on $V_2$, we have introduced edges $\{a1\}$, and $\{b2, b3, b4\}$. \\

\begin{figure}[h!]
	\centering
	\includegraphics[scale=0.6]{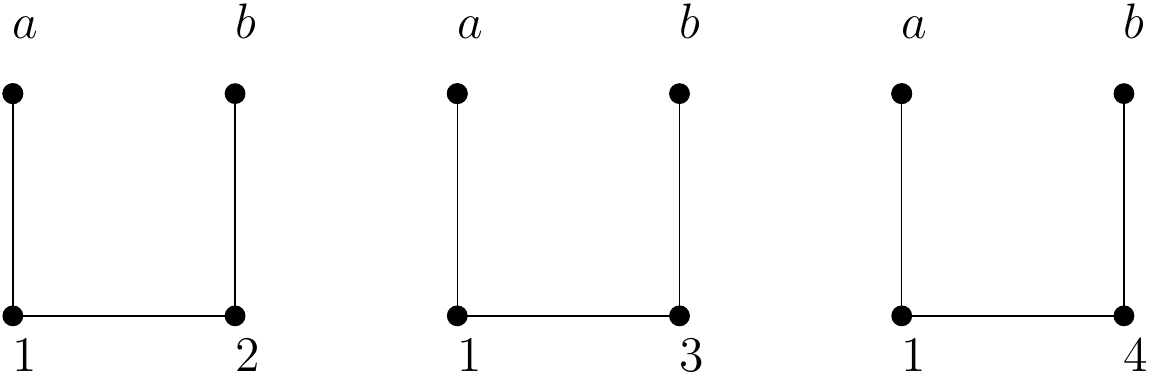}
	\caption{\textit{Second step in generating an unswitchable graph}}
	\label{uggTwo}
\end{figure}

For each one of the edges of the clique we consider the induced $P_4$ formed with pairs of vertices in the set $V_1 = \{a,b\}$.
There are three of them as shown in Figure~\ref{uggTwo}.\\

\begin{figure}[h!]
	\centering
	\includegraphics[scale=0.6]{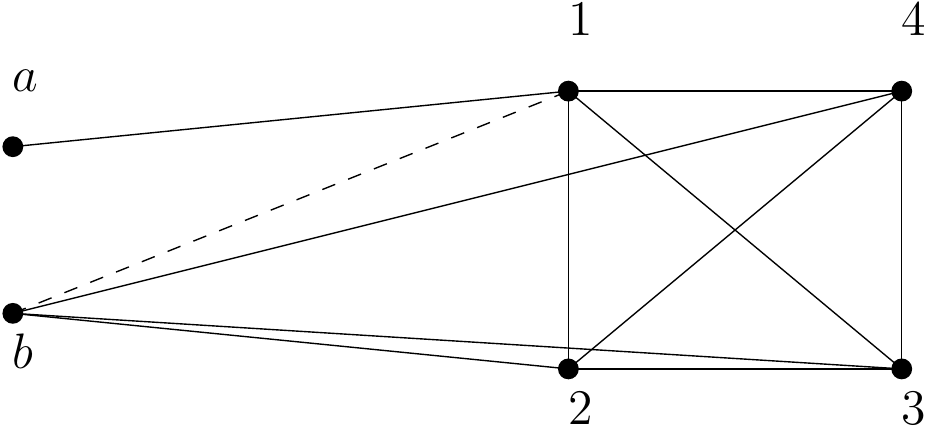}
	\caption{\textit{Third step in generating an unswitchable graph}}
	\label{uggThree}
\end{figure}

These induced subgraphs can be taken care of by introducing one the edges in $\{a2, b1\}$, $\{a3, b1\}$ and $\{a2, b1\}$ in the induced  $P_4$'s from left to right. 
All three can be taken care of by introducing the edge 
$b1$ in the three induced $P_4$'s. The updated graph is shown in Figure~\ref{uggThree}, with the newly added edge as a dashed segment.\\

We immediately see that this as a problem of finding minimum cover for a class of 2-element sets in the general case. 
We have to go further. \\

Introducing these new edges can give rise to new induced $P_4$'s. These are found by examining pairs
of newly introduced edges and checking whether a $P_4$ is induced by these edges and the edge joining their end points in the 
set $V_2$. Once again, we generate a class of 2-element sets, for which we solve a minimum cover problem. We continue this  
iteratively, until we reach a stage when we have a cover of size one. \\

In the chosen example above, the process comes to an end in one step.  \\

We describe formally the algorithms to generate an instance of the vertex cover problem
introduced at each stage and since it is an NP-complete problem a minimum-vertex degree 
heuristic used to add as few chords as possible to eliminate the $P_4$'s. \\

\begin{algorithm}
	\caption{\textit{vertexCoverInstanceGeneration}}
	\begin{algorithmic}[1]
		\State \textbf{Input:} $ListP4$ from $G$
		\State \textbf{Output:} An ordered dictionary with the two vertices that define a chord as key and value the frequency of occurrences of the chord.  
		\Procedure{find\_all\_2\_element\_sets}{$ListP4$}
		\State $two\_element\_set\_count \leftarrow \{\}$
		\For {each $e \in ListP4$}
		\State $two\_element\_inst\_one \leftarrow$ first and third element of $e$
		\State $two\_element\_inst\_two \leftarrow$ second and fourth element of $e$
		\If{$two\_element\_inst\_one$ is present in $two\_element\_set\_count$}
		\State {Increment the value by 1 for key $two\_element\_inst\_one$}
		\Else{}
		\State {Set the value to 1 for key $two\_element\_inst\_one$}
		\EndIf
		\If{$two\_element\_inst\_two$ is present in $two\_element\_set\_count$}
		\State {Increment the value by 1 for key $two\_element\_inst\_two$}
		\Else{}
		\State {Set the value to 1 for key $two\_element\_inst\_two$}
		\EndIf
		\EndFor
		\State {Sort  $two\_element\_set\_count$ in decreasing order of values}
		\State \textbf{return} $two\_element\_set\_count$
		\EndProcedure
	\end{algorithmic}
\end{algorithm}

\newpage
\begin{algorithm}
	\caption{\textit{edgeAddition}}
	\begin{algorithmic}[1]
		\State \textbf{Input:} Current edge list of graph $G$, $ListP4$ and $V_1$
		\State \textbf{Output:} List of new edges added to make $G$ unswitchable.  
		\Procedure{add\_min\_edges}{$edge\_list$, $ListP4$, $V_1$}
		\State $edge\_list\_copy \leftarrow$ Copy current $edge\_list$
		\State $new\_edge\_list \leftarrow []$
		\While{$ListP4$ is not empty}
		\State $two\_element\_counts \leftarrow$  find\_all\_2\_element\_sets($ListP4$)
		\State $vertices\_to\_add \leftarrow$ first element of $two\_element\_counts$
		\State Add the extracted vertices from $vertices\_to\_add$ to $edge\_list\_copy$ 
		\State Add the extracted vertices from $vertices\_to\_add$ to $new\_edge\_list$ 
		\State $adjacency\_list \leftarrow$ Convert $edge\_list\_copy$ to adjacency list
		\State $ListP4 \leftarrow$ find\_all\_p4s($adjacency\_list$, $V_1$)
		\EndWhile
		\State \textbf{return} $new\_edge\_list$
		\EndProcedure
	\end{algorithmic}
\end{algorithm}

Finally, we put everything together and describe formally our algorithm for generating 
an unswitchable graph. \\

\begin{algorithm}
	\caption{\textit{unswitchableGraphGeneration}}
	\begin{algorithmic}[1]
		\State \textbf{Input:} Number of vertices, $n_1$, in independent set $V_1$ and number of vertices, $n_2$, in clique set $V_2$
		\State \textbf{Output:} An edge\_list describing the graph $G$.  
		\Procedure{generate\_unswitchable\_graph\_for}{$n_1$, $n_2$}
		\State $edge\_list \leftarrow []$
		\State Add edges corresponding to the vertices in clique set to edge\_list
		\State $count\_in\_V_1 \leftarrow$ Pick random number of vertices in $V_1$ to connect to $V_2$
		\While{$count\_in\_V_1$}
		\State $vertex\_V_1\_index \leftarrow$  Pick a random vertex from $V_1$
		\State $vertices\_count\_to\_connect \leftarrow$ Pick a random number of vertices of $V_2$ to connect
		\While{$vertices\_count\_to\_connect$}
		\State $vertex\_V_2\_index \leftarrow$ Pick a random vertex in clique set
		\State $edge \leftarrow$ ($vertex\_V_1\_index$, $vertex\_V_2\_index$)
		\If{$edge$ not in edge\_list}
		\State Add the edge to $edge\_list$ and decrement $vertices\_count\_to\_connect$
		\EndIf
		\EndWhile
		\State Decrement $count\_in\_V_1$
		\EndWhile
		\State $adjacency\_list \leftarrow$ Convert $edge\_list$ to adjacency list
		\State $new\_edge\_list \leftarrow []$
		\State $ListP4 \leftarrow$ find\_all\_p4s($adjacency\_list$, $V_1$)
		\If {number of $p4s > 0$}
		\State $new\_edge\_list \leftarrow$ add\_min\_edges($edge\_list$, $ListP4$, $V_1$)
		\EndIf
		\State Concatenate $new\_edge\_list$ to $edge\_list$
		\State \textbf{return} $edge\_list$ that corresponds to unswitchable graph $G$
		\EndProcedure
	\end{algorithmic}
\end{algorithm}

Now that we have discussed a method for generating unswitchable graphs, it is instructive to choose an $n$ and construct sets $S_1, S_2, \ldots, S_n$, distributing the vertices of the graph 
in these sets so that the adjacencies are exactly the same as in the graph of Figure~\ref{uggThree}. \\
 
\begin{figure}[h!]
	\centering
	\includegraphics[scale=0.6]{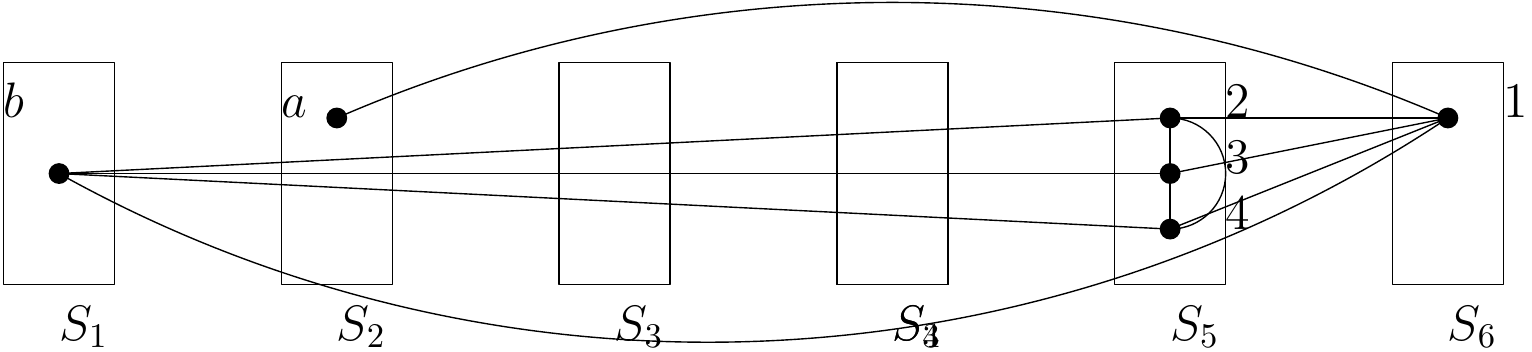}
	\caption{\textit{Construction of the graph of Figure~\ref{uggThree} by applying Eggleton's theorem}}
	\label{uggFour}
\end{figure}

Set $n = 6$ and define the sets
$S_i$ as follows: $S_1 = \{b\}$, $S_2 = \{a\}$, $S_3 = S_4 = \{ \}$, $S_5 = \{2, 3, 4\}$ and $S_6 = \{1\}$. From Eggleton's
theorem the adjacencies of the vertices in these sets are as shown in Figure~\ref{uggFour} and we have the same graph as in Figure~\ref{uggThree}. \\

Let $N = \#P_4$ the number of $P_4$'s discovered in the first step of the vertex cover algorithm. 
The time complexity of the generation algorithm is then $O(n^3 + N^2)$, where the term $n^3$, as in the 
recognition algorithm, accounts for the time complexity of identifying the $P_4$'s and the second term is the
sum obtained by adding a sequence of $P_4$'s starting with $N$ and decreasing to one, each term being an upper bound
on the size of the vertex cover problem to be solved.

\section{Conclusions}
In this note we have proposed an algorithm for recognizing unswitchable graphs, by first showing that unswitchable
graphs are a subclass of split graphs. The second half of the proof of Eggleton's theorem requires an unswitchable graph as input. Motivated by this, we have proposed an interesting algorithm for generating unswitchable graphs. The third author has implemented both the algorithms in Python 3. In the light of Theorem 2, the degree sequence of an unswitchable graph is uniquely
realizable. \\

 A challenging open problem is to design an algorithm for generating an unswitchable graph on $n$ vertices uniformly at random.

\end{document}